\documentclass[preprint2]{aastex}
\usepackage{psfig}

\newcommand{\WV}{$^{185}$W}
\newcommand{\WVI}{$^{186}$W}
\newcommand{\rng}{($n$,$\gamma$)}
\newcommand{\rgn}{($\gamma$,$n$)}

\shorttitle{The $s$-process branching at $^{185}$W}
\shortauthors{Sonnabend et al.}

\begin{document}


\title{The $s$-process branching at $^{185}$W}


\author{K.~Sonnabend, P.~Mohr, K.~Vogt, and A.~Zilges}
\affil{Institut f\"ur Kernphysik, Technische Universit\"at Darmstadt, \\
Schlossgartenstra{\ss}e 9, D-64289 Darmstadt, Germany}
\author{A.~Mengoni}
\affil{ENEA, V.le G.~B.~Ercolani 8, \\
        I-40138 Bologna, Italy}
\author{T.~Rauscher}
\affil{Institut f\"ur Physik, Universit\"at Basel, \\
  Klingelbergstrasse 82, CH-4056 Basel, Switzerland
}
\author{H.~Beer and F.~K\"appeler}
\affil{Forschungszentrum Karlsruhe, Institut f\"ur Kernphysik, \\
P.O. Box 3640, D-76021 Karlsruhe, Germany}
\and
\author{R.~Gallino\footnote{temporary address: Max-Planck-Institut f\"ur
    Chemie, Abteilung Kosmochemie, Becherweg 27, D-55128 Mainz, Germany}}
\affil{Dipartimento di Fisica Generale, Universit{\'a} di Torino \\
and Sezione INFN di Torino, Via P.~Giuria 1, I-10125 Torino, Italy}


\begin{abstract}
The neutron capture cross section of the unstable nucleus \WV\ has
been derived from experimental photoactivation data of the inverse
reaction \WVI \rgn \WV . The new result of
$\sigma =  (687 \pm 110)$\,mbarn confirms the
theoretically predicted neutron capture cross section of \WV\ of
$\sigma \approx 700$\,mbarn at $kT = 30$\,keV. A neutron density
in the classical $s$-process of $n_{\rm n} = (3.8 ^{+0.9} _{-0.8}) \times 
10^8$ cm$^{-3}$ is derived from the new data for the \WV\ branching.
In a stellar $s$-process model one finds a significant overproduction
of the residual $s$-only nucleus $^{186}$Os.
\end{abstract}




\section{Introduction}
\label{sec:intro}
The unstable nucleus \WV\ is a so-called branching point in the slow
neutron capture process ($s$-process). The nucleus \WV\ is produced by neutron
capture in the $s$-process from the stable $^{184}$W. At small neutron
densities \WV\ $\beta$-decays to $^{185}$Re with a half-life of
$T_{1/2} = 75.1$\,d, 
and it has been pointed out that the $\beta$-decay half-life does
practically not depend on the temperature at typical $s$-process
conditions \citep{Taka87}. At higher neutron densities \WV\ may
capture one more neutron leading to the stable \WVI . It is obvious
that the branching between $\beta$-decay and neutron capture depends
on the $\beta$-decay half-life, the neutron capture cross section, and
the neutron density. The half-life and the neutron capture cross
section can be measured in the laboratory, and therefore one can
determine the neutron density from the observed abundances of the
various tungsten isotopes \citep{Kaepp91}. Additionally, this
branching has minor influence on the $^{187}$Os/$^{187}$Re
cosmochronometer \citep{Bosch96}.

Up to now, only theoretical estimates are available for the neutron
capture cross section of \WV\ because direct neutron capture
experiments with radioactive targets are very difficult. Theoretical
predictions for the Maxwellian averaged capture cross section at a
typical temperature of 30\,keV vary significantly from 532\,mbarn
\citep{Kaepp91} and 560\,mbarn \citep{Rau00} to 794\,mbarn \citep{Hol76}. In
a recent compilation a value of $(703 \pm 113)$\,mbarn has been adopted
\citep{Bao00}. All calculations used the statistical model. The
differences in the results come from the parameterizations of the level
density, the gamma-ray strength function, and the neutron-nucleus
optical potential. 

In order to reduce the uncertainties, a new experiment was performed
on the inverse reaction \WVI \rgn \WV . The idea is to find a
parameter set for the calculations which reproduces the cross section
of \WVI \rgn \WV , and to apply these parameters for the prediction of
the \WV \rng \WVI\ cross section. Such a prediction should be more
reliable for one special reaction than previous calculations which
used global or local systematics to derive the relevant parameters
from neighboring nuclei. The relevant energy region is located close
above the threshold of the \rgn\ reaction at $S_n = 7194$\,keV
\citep{Mohr01}. At higher energies experimental data on the \WVI \rgn
\WV\ reaction are available in literature \citep{Ber69,Gor78,Gur81},
and the results can be found in the compilations of \citet{Die88} and
in CDFE \citep{CDFE}. We have performed an
additional measurement at energies close above the threshold.

In \S\,\ref{sec:exp} we present our experimental set-up. 
In \S\,\ref{sec:calc} we calculate the cross sections of the \rng\ and
\rgn\ reactions, in \S\,\ref{sec:branch} we derive the $s$-process
neutron density from our experimental data, and we apply a stellar
$s$-process model to the \WV\ branching. \S\,5 gives a summary
and conclusions.

\section{Experimental Set-up and Procedure}
\label{sec:exp}
The \WVI \rgn \WV\ experiment was performed using the photoactivation
technique at 
the real photon set-up at the superconducting linear electron
accelerator S-DALINAC \citep{Rich96}. Recently, several
photoactivation experiments have been performed here
\citep{Mohr00,Vogt01,Lind01,Mohr00b}, and \rgn\ cross sections and
reaction rates were determined using a quasi-thermal photon bath at
temperatures of several $10^9$\,K. These data are relevant for the
nucleosynthesis of the neutron-deficient so-called $p$-nuclei
\citep{Lam92}.

\begin{figure}
  \begin{center}
    \psfig{figure=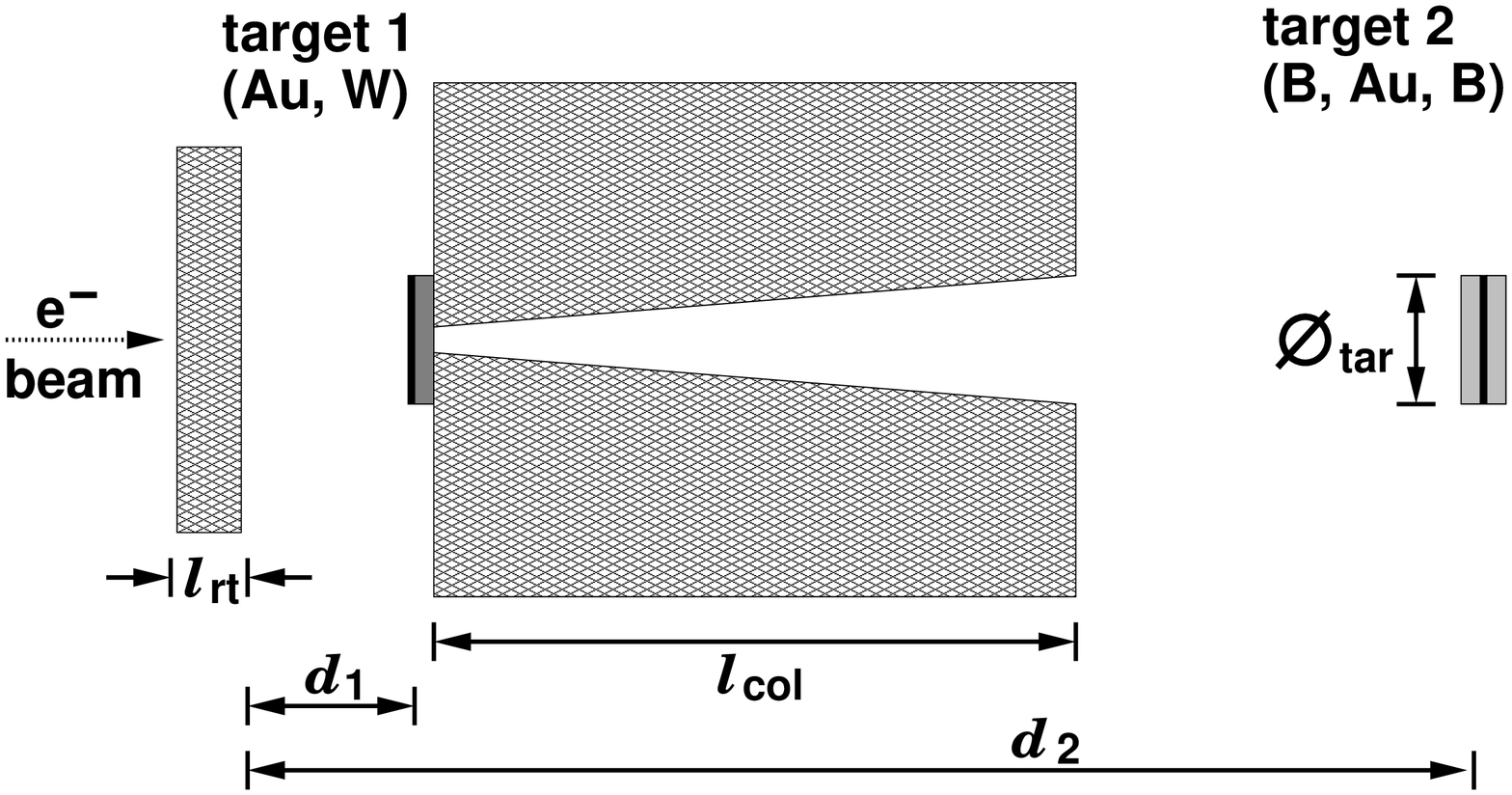,width=7.5cm}
  \end{center}
  \caption{
    Scheme of the photoactivation setup at the S--DALINAC. The
    radiator target ($l_{\rm rt}$~=~1.4~cm) which generates
    the photon beam and the collimator ($l_{\rm col}$~=~95.5~cm) are
    sketched as well as the irradition positions of our targets
    ($d_1 \approx$~5~cm , $d_2 \approx$~150~cm,
    $\varnothing_{\rm tar}$~=~2~cm). Note that the lengths are off scale!
    \label{plotone}
    }
\end{figure}

Photons were generated by bremsstrahlung using our electron beam at an
energy of $E_0 = 8775$\,keV and with a beam current of about
$30$\,$\mu$A. Usually, the photon beam is collimated and hits the
target at a distance $d_2 \approx$~150\,cm behind the radiator
target. This leads to a well-defined photon beam with a spectral
composition which was analyzed in detail \citep{Vogt01}. However,
because of the relatively long half-life of \WV\ ($T_{1/2} = 
(75.1 \pm 0.3)$\,days) and the weak $\gamma$-ray branch ($E_\gamma =
125.4$\,keV) in the $\beta$-decay of \WV\ to $^{185}$Re of only $(1.92
\pm 0.07) \times 10^{-4}$ \citep{Kem92}
the irradiation of \WVI\ had to be performed with the highest photon
flux that can be obtained at our irradiation set-up. Therefore, the
tungsten target was mounted directly behind the radiator target
($d_1 \approx$~5\,cm) where
the photon intensity is roughly a factor of 300 higher than at our
usual irradiation position. Fig.~\ref{plotone} shows an overview of
our experimental setup. The target consisted of a thin metallic
tungsten disk of natural isotopic composition with a diameter of
$\varnothing =$~20\,mm and a thickness of about 1.2\,mm. Note that the
amount of target material remains limited because of the absorption of the
low-energy decay $\gamma$-ray in the target. Properties of the target
are summarized in Table \ref{tableone}. The decay properties of the
residual nuclei are listed in Table \ref{tabletwo}.

To minimize systematic uncertainties from the photon flux
determination at the position directly behind the radiator target, a
relative measurement was carried out. We irradiated simultaneously
the tungsten target and a very thin (25\,$\mu$m) gold disk at the
position close to the radiator target. A second thin gold disk was
sandwiched between two layers of boron. This sandwich target was
mounted at the regular target position and irradiated simultaneously
with the tungsten and gold targets close to the radiator. The boron
target is used to normalize the incoming photon intensity by the
$^{11}$B($\gamma$,$\gamma'$) reaction. From the absolute photon
intensity at the regular target position and from the activation of
the second thin gold target one can determine the \rgn\ cross section
of $^{197}$Au. The complete determination of the $^{197}$Au\rgn
$^{196}$Au cross section is presented in \citet{Vogt02}. And, finally,
the \WVI \rgn\ cross section can be determined from the ratios of
activities of the gold and tungsten targets close to the radiator. 
Absorption of bremsstrahlung $\gamma$-rays in the targets can be
neglected. Typical uncertainties for the photon flux determination are
of the order of 10\,\%. Further details of the experimental set-up can
be found in \citet{Vogt01} and \citet{Mohr99}. 

\begin{figure}
  \begin{center}
    \psfig{figure=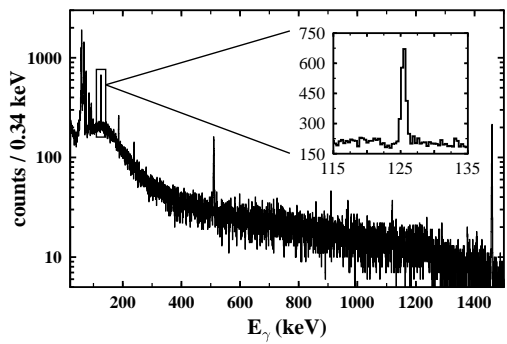,width=7.5cm}
           %bbllx=90pt,bblly=55pt,bburx=685pt,bbury=490pt,width=15.0cm}
  \end{center}
  \caption{
    Typical $\gamma$-ray spectrum of the activated tungsten target. The
    125.4\,keV line from the weak $\gamma$-ray branching in the decay
    \WV\ $\rightarrow$ $^{185}$Re can clearly be identified (see
    inset). The measuring time for this spectrum was twelve hours.
    \label{plottwo}
    }
\end{figure}

The decay $\gamma$-rays of the activated tungsten and gold targets
were measured using a 
well-shielded high-purity germanium (HPGe) detector with a relative
efficiency of 30\,\% and an energy resolution of 2\,keV (at
1332.5\,keV). A typical spectrum of the tungsten target is shown in
Fig.~\ref{plottwo}. We followed the decay of the activity over more
than one half-life, and the analysis of our decay curve leads to a
half-life of $T_{1/2} = (76.6 \pm 1.5)$\,d which agrees with the adopted
value of $T_{1/2} = (75.1 \pm 0.3)$\,d \citep{ensdf} within the uncertainties.
The precise exponential decay of the activity 
confirms that the analyzed $\gamma$-ray line does not accidentally
overlap with a background line. Of course, also the excellent energy
resolution of the HPGe detector helps to measure a weak $\gamma$-ray
branching. The efficiency of the HPGe detector was determined by
calibrated sources \citep{Vogt01}. Additionally, for the 125.4\,keV
$\gamma$-ray from the \WV\ decay the self-absorption in the tungsten
target was taken into account by GEANT simulations \citep{geant}, and the GEANT
simulations of the absorption in tungsten were verified by transmission
measurements of the tungsten target. The resulting uncertainty of the
relative efficiency is about 5\,\%.

The yield $Y$ in our experiment is proportional to the
energy-integrated cross section $I_{\sigma}$:
\begin{equation}
Y \sim I_\sigma = \int_{S_n}^{E_0} N_{\rm{br}}(E,E_0) \, \sigma(E) \, dE
\label{eq:yield}
\end{equation}
with $E_0$: endpoint energy of the bremsstrahlung and kinetic energy
of the electron beam; $N_{\rm{br}}(E,E_0)$: number of bremsstrahlung
photons at an energy $E$ and with the endpoint energy $E_0$;
$\sigma(E)$: \WVI \rgn \WV\ cross section. The factor between the
yield $Y$ and the energy-integrated cross section $I_{\sigma}$ depends
on characteristics of the observed isotope as well as on parameters of
the experimental setup \citep{Vogt01}. From the experimentally
measured yield ratio $Y_{\rm{W}}/Y_{\rm{Au}}$ between the tungsten and
the gold yields one can derive the ratio of
the integrated cross sections
$I_{\sigma,{\rm{W}}}/I_{\sigma,{\rm{Au}}} = 12.3 \pm 0.9$
with relatively small uncertainties. Note that the relatively large
value of this ratio does not indicate that the cross section of
\WVI\ is much larger than the cross section of $^{197}$Au. The reason
for this large ratio is the much smaller neutron separation energy
$S_n$ of \WVI\ compared to $^{197}$Au which leads to a broader
integration range in Eq.~(\ref{eq:yield}) for \WVI .

It is not possible to determine the energy dependence of the \WVI \rgn
\WV\ cross section from one photoactivation measurement with a white
bremsstrahlung spectrum. If one adopts the
theoretically calculated energy dependence of the cross section
$\sigma(E)$ (see \S\,\ref{sec:calc}), it is possible to solve the
integral in Eq.~(\ref{eq:yield}) and to determine a normalization
factor $F$ for the theoretical calculation by comparison of the
theoretically predicted and experimentally measured yields. In the
case of the \WVI \rgn \WV\ reaction this leads to normalization
factors of $F = Y_{\rm{exp}}/Y_{\rm{calc}}$ close to unity for two
different calculations of the \WVI \rgn \WV\ cross section (see
\S\,\ref{sec:calc}). 
The same factors $F$ should be used for the prediction of the inverse
\WV \rng \WVI\ cross section. 

The reaction $^{197}$Au\rgn $^{196}$Au has been used to normalize the
\WVI \rgn \WV\ experiment.  The determination of the $^{197}$Au\rgn
$^{196}$Au cross section has been performed similar to \citet{Vogt01}
and is published elsewhere \citep{Vogt02}. Our new data agree nicely with
several previous experiments \citep{Ber87,Vey70} and a recent
experiment with monochromatic photons from Laser-Compton
backscattering \citep{Uts01}.

The experimental uncertainties are dominated ($i$) by the photon flux
determination which enters into the analysis of the experimental
yields in Eq.~(\ref{eq:yield}) for both the target \WVI\ and the
standard $^{197}$Au, and ($ii$) by the self-absorption of the
125.4\,keV $\gamma$-ray 
in the tungsten target. The total uncertainty for the normalization
factors $F$ is 14\,\%.

\begin{figure}
  \begin{center}
    \psfig{figure=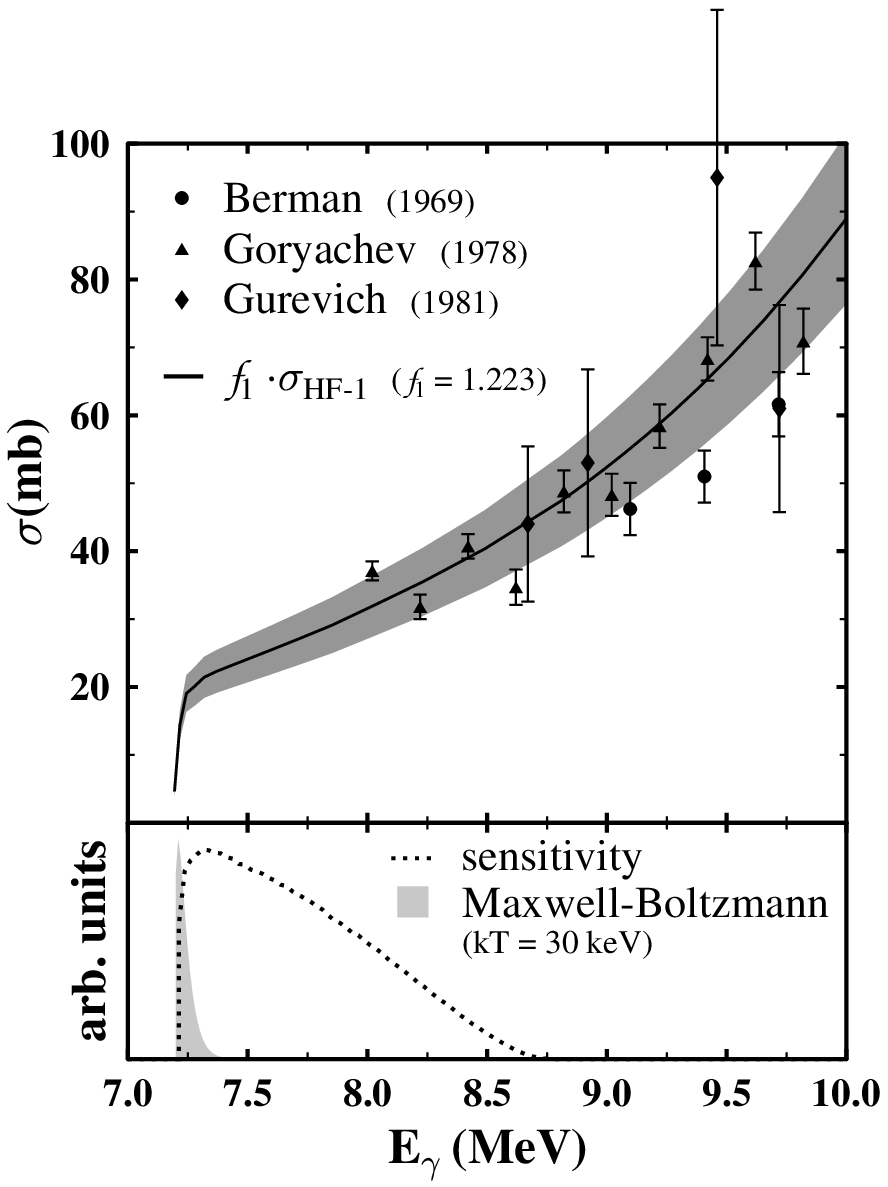,%height=15.5cm}
bbllx=0pt,bblly=0pt,bburx=257pt,bbury=315pt,width=7.5cm,clip=}
  \end{center}
  \caption{
    Cross section of the \WVI \rgn \WV\ reaction. At higher energies data
    from literature are available \citep{Ber69,Gor78,Gur81}. 
    The theoretical calculation HF-1 (see Sec.~\ref{sec:calc}),
    normalized by $F_{1} = 1.223$ 
    (upper part, full line, uncertainties gray shaded) 
    agrees nicely with these data. The sensitive energy range of
    our new experiment is located close above the \rgn\ threshold
    (lower part, dotted line). The astrophysically relevant energy
    range which is defined by a Maxwell-Boltzmann distribution with
    $kT = 30$\,keV above the threshold at 7194\,keV is also shaded
    (lower part, light gray).
    \label{plotthree}
    }
\end{figure}

\begin{figure}
  \begin{center}
    \psfig{figure=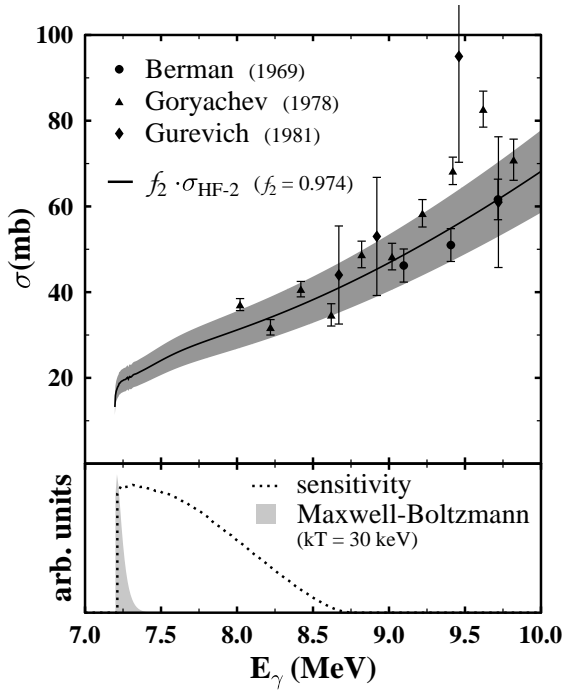,%height=15.5cm}
bbllx=0pt,bblly=0pt,bburx=257pt,bbury=315pt,width=7.5cm,clip=}
  \end{center}
  \caption{
    Cross section of the \WVI \rgn \WV\ reaction. 
    Same as Fig.~\ref{plotthree}, but with the
    theoretical calculation HF-2 (see Sec.~\ref{sec:calc}),
    normalized by $F_{2} = 0.974$ 
    (upper part, full line, uncertainties gray shaded) which again
    agrees nicely with the available data. The sensitive energy range of
    our new experiment is located close above the \rgn\ threshold
    (lower part, dotted line). The astrophysically relevant energy
    range which is defined by a Maxwell-Boltzmann distribution with
    $kT = 30$\,keV above the threshold at 7194\,keV is also shaded
    (lower part, light gray).
    \label{plotfour}
    }
\end{figure}

The normalized calculations are compared with 
experimental data at higher energies in Figs.~\ref{plotthree} and
\ref{plotfour}. 
Within the uncertainties one finds excellent agreement. Additionally,
the dotted line shows the integrand of Eq.~(\ref{eq:yield}); this line
defines the energy range where our experiment was sensitive. As a
consequence of the white bremsstrahlung spectrum and the calculated
energy dependence of the \rgn\ cross section, the energy range of
our experiment is located directly above the \rgn\ reaction threshold,
and it has a width of roughly 1\,MeV. It is not possible to
present our result as one data point in Figs.~\ref{plotthree} and
\ref{plotfour}. 
Instead, the experimental result of this work are normalization
factors $F$ for the theoretical calculations.

\section{Calculation of the \rgn\ and \rng\ Cross Sections}
\label{sec:calc}

Two sets of calculations, HF-1 and HF-2, 
of the relevant cross sections have been performed. 
In both cases the statistical model
Hauser-Feshbach theory has been applied to describe
the reaction process. Then, in one case, HF-1, a global set
of parametrization was used, where the main model parameters
were derived either from microscopic approaches
or from global systematics.
In the other case, HF-2, the model parameters were optimized 
to the mass region under consideration and, when possible, 
parameters derived from experimental nuclear structure data 
were used.

The calculation HF-1 was performed with 
the code NON-SMOKER \citep{Rau98,Rau00}.
The neutron transmission coefficients were computed 
using a microscopic potential \citep{Jeu77}.
Nuclear levels as given in \citet{Rau01} have been utilized. 
Above the last known state a global theoretical level density 
description was used \citep{Rau97}. 
The E1 $\gamma$-transition probabilities were described 
by a Lorentzian shape with a  modified low-energy tail,
following the prescription of \citet{Cul81}. 
Width and energy position of the GDR were also taken from theory. 
From a hydrodynamic droplet approach~\citep{Mye77} 
we obtain $E_{\rm GDR}=14.23$ MeV, and from a parameterized 
approach \citep{Cow91} the width is determined as
$\Gamma_{\rm GDR}=5.45$ MeV. The modified energy-dependent width is then 
given as $\Gamma (E_\gamma)=\Gamma_{\rm GDR}\sqrt{E_\gamma/E_{\rm GDR}}$. 
For a deformed nucleus, the energy and width split %become split 
according to the description outlined in \citet{Cow91}.
However, within the droplet model \citep{Mye77} 
the nucleus $^{186}$W is spherical and a single-humped Lorentzian
with the above energy and width is obtained.

The calculation HF-2 was performed using the optical model
parameters (OMP) of \citet{Mol65}, 
for neutron transmission coefficients.
This set of OMP reproduces fairly well the total cross sections of nuclei
with $A$~=~184-188. For example, the total cross section of 
$^{186}$W is reproduced with an accuracy of better than 10\%
for neutron energies from 100\,keV up to 10\,MeV, by this OMP set.
Gamma-ray transmission coefficients were derived
from the experimental double-humped GDR parameters~\citep{Die88}
derived from experimental $(\gamma,n)$ data in the GDR region.
The relevant data are: 
$E_{\rm GDR} = 12.59$\,MeV and 14.88\,MeV, 
$\Gamma_{\rm GDR} = 2.29$\,MeV and 5.18\,MeV, and
$\sigma^{\rm peak}_{\rm GDR} = 211$\,mbarn and 334\,mbarn,
for the two GDR components.
Nuclear level densities were derived from the parametrization
of \citet{Men94}. Experimental discrete levels
have been used to fit the constant-temperature parametrization
at low excitation energies, matched to the pairing+shell
corrected Fermi-gas model (Gilbert-Cameron prescriptions)
at excitation energies close to the neutron binding energy.

In order to compare the results of model predictions with
the present $^{186}$W$(\gamma,n)^{185}$W experimental data, 
the calculation of the cross sections for $^{185}$W 
in the ground state, as well as in
several excited states have to be performed.
Here, we have included excited states up to 
about 500\,keV. Higher excited states do not
appreciably contribute to the cross section.
This request is altogether similar to what is
needed to evaluate the $(n,\gamma)$ cross section 
for thermally excited target states in
stellar plasma (stellar cross sections).

From a comparison with the present experimental data 
close to the neutron threshold, a renormalization factor 
$F_{1}=1.223$ is obtained for the global HF-1, 
and $F_{2}=0.974$ for the local HF-2 calculation.
Both calculations are able to reproduce the measured data from the
neutron threshold up to the GDR region (see Figs.~\ref{plotthree}
and \ref{plotfour}).
The typical uncertainty of 25 to 30 \% associated with 
neutron capture cross section calculations %of the order of 25 to 30 \%  
(see for example the prediction of the NON-SMOKER code \citep{Rau97,Bao00})
is therefore obtained with either model parameterizations.

It is interesting to note the similarity in the results of both
calculations despite the strongly differing treatments of the GDR, which
would be expected to dominate the difference in the results. Although
HF-1 uses a single-humped GDR shape, a similar strength
distribution as for HF-2 with its double-humped GDR
is obtained at the low-energy side of the GDR.
This is due to the use of an energy-dependent width, a slightly smaller
GDR energy, and a slightly broader basic GDR width. Thus, the
energy-dependence of the resulting cross section in the relevant energy
range is almost the same for both calculations.
The absolute values are roughly proportional to the $\gamma$-ray
strength function, hence, they show a comparatively small difference.

The impact of other model inputs, such as the optical neutron potential,
%and the nuclear level density, 
are rather small. When comparing the
Maxwellian averaged cross sections obtained in the two description, we
notice increasing deviations at the lowest energies, i.e.\ below 10 keV.
This means that a different energy dependence of the cross sections is
found, being mainly due to the different optical model potentials used.
At higher energies the energy dependence of the HF-1 and HF-2
calculations is similar and thus the uncertainties stemming from the
optical potentials are not significant to explain further differences
in the two approaches.

We will assume that the same renormalization factors 
derived from the present experimental data and
model calculations, are valid also for the Maxwellian averaged 
cross section of the reverse reaction  
\WV \rng \WVI . %Strictly speaking, this is a
%simplification because the bremsstrahlung spectrum used in the
%measurement is not a thermal one, thus, detailed balance does not apply. 
%However, we argue that the largest uncertainties in the present
%calculations stem from the model parameters and that the uncertainty
%associated with these inputs is larger than the error introduced by the
%assumption of a global renormalization factor.
This assumption requires some more discussion.
What is actually measured and calculated here is a compound reaction in
which the ``compound'' nucleus consists of excited states of $^{186}$W created
by $\gamma$-excitation of the ground state of $^{186}$W. This compound
nucleus subsequently decays in the neutron channel to all energetically
possible final states in $^{185}$W, according to the Bohr hypothesis,
i.e.\ independently of how it was formed. For a full application of detailed
balance linking the \rng\ and \rgn\ reaction rates or
Maxwellian averaged cross sections, one would have to use a Planck
distribution of photons according to the astrophysical temperature of
interest ($\simeq 0.348\times 10^9$ K) plus account for thermal
excitation of the target at the same temperature instead of keeping the
target in the ground state. Thus, for each photon energy we are actually
measuring a subset of the transitions relevant for the capture rate.
Applying detailed balance directly yields a neutron capture cross
section which is the thermally averaged sum of neutron captures on the
ground state and excited states of $^{185}$W, forming $^{186}$W at the
given energy and finally directly decaying to the ground state of the
final nucleus $^{186}$W. However, this cross section is governed by the
same uncertainties in the optical neutron potential, the level
densities, and the GDR properties as the full stellar cross section.
Since the energy dependence of the \rgn\ reaction is well
described by both models, it can safely be assumed that there is no
further energy dependence in the renormalization factor.
Therefore, we argue that the same renormalization factor can be applied
also to the Maxwellian averaged cross section at 30 keV.

\subsection{Results for \WV \rng \WVI }
\label{subsec:results}
The normalized results of both calculations for the Maxwellian averaged 
cross section of \WV \rng
\WVI\ at $kT$=30\,keV are in remarkable agreement: from HF-1 one
obtains $\sigma = 734$\,mbarn, and from HF-2 $\sigma = 640$\,mbarn.
The average value is
687\,mbarn $\pm 100$\,mbarn (experimental uncertainty) $\pm 47$\,mbarn from
different calculations leading to a final result of $(687 \pm
110)$\,mbarn. This value is already the stellar capture cross section
at $kT$~=~30 keV where an enhancement factor of 0.92 was used; the cross
section at $kT \approx 0$ is 747\,mbarn which is in agreement with the 
adopted value of $(703 \pm 103)$\,mbarn \citep{Bao00} 
within the uncertainties. %Contrary to the previously adopted
%value which was only based on theoretical calculations, the new result
%is based on experimental data of the inverse reaction.
Contrary to this adopted value which was based on an empirical
renormalization of a theoretical value, the new result is based on
experimental data of the inverse reaction. The compatibility of both
values nicely confirms the estimate of the systematic error in the
theoretical calculation which was used to generate the adopted value in
\citet{Bao00}.

\section{The Branching at \WV }
\label{sec:branch}
\subsection{The classical $s$-process}
\label{subsec:classical}
\begin{figure}
  \psfig{figure=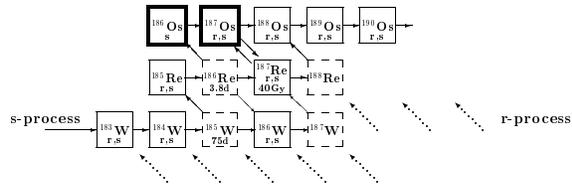,width=7.5cm}
  \caption{\label{plotfive} The $s$-process path in the W-Re-Os mass 
    region. Two branchings occur at \WV\ and at $^{186}$Re. Unstable
    nuclei are marked by dashed boxes (except the quasi-stable
    $^{187}$Re). The indicated values are the terrestrial
    half-lives. Note that the half-life of $^{187}$Re decreases by ten
    orders of magnitude at stellar temperatures and common densities
    of the He-intershell while $^{187}$Os becomes unstable
    \citep{Taka87}.
    }
\end{figure}

The $s$-process synthesis path in the W-Re-Os mass region is shown in
Fig.~\ref{plotfive}. The nuclei $^{186,187}$Os are partially bypassed
by the $s$-process flow especially by a branching at $^{185}$W. The
stellar beta decay rates of $^{185}$W and $^{186}$Re are practically
independent of temperature in the relevant temperature region
\citep{Taka87}. Therefore, the isotopic abundance of $^{186}$Os is
determined by the average $s$-process neutron density. In
Fig.~\ref{plotfive} laboratory half-lives are indicated. However,
according to \citet{Taka87}, at stellar temperatures of interest the
${187}$Re nucleus is almost fully ionized and its $\beta^-$-decay rate
increases by about ten orders of magnitude ($T_{1/2}$ = 21.3 yr at a
temperature $T$ = 3 $\times$ 10$^8$ K and electron density $n_e$ =
10). In the same stellar envrironment, ${187}$Os, which is stable in
terrestrial conditions, becomes unstable by electron capture
($T_{1/2}$ = 1243 yr). Consequently, the abundance ratio between the
two nuclei in the He intershell production zone and in the AGB
envelope needs to be followed carefully. 
  
The classical $s$-process model formulated by \citet{wnc76} provides 
a simple way of estimating the $s$-process neutron density 
via branching analyses. Previous analyses
\citep{Kaepp91} of the W-Re-Os branching can now be updated 
with a neutron capture cross section for $^{185}$W 
which is not only based on statistical model calculations. 

\begin{figure}
  \psfig{figure=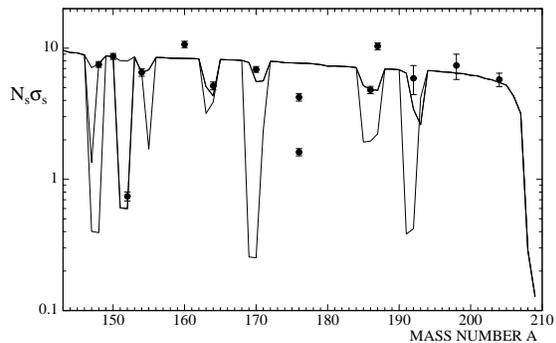,width=7.5cm}
  \caption{\label{plotsix} The calculated $N_s\sigma_s$ curve (solid 
    line) shown in the mass range from $A=145$ to the termination of the 
    $s$-process at $^{209}$Bi. Empirical data are given as solid circles. 
    Important branchings of the $s$-process path occur at 
    $A=147-148$, $151-152$, $154-155$, $163-164$, $169-171$, $185-187$,
    and $191-192$. 
    The curve is normalized to $^{150}$Sm on the unique path.
    The data points at $A=176$ and 187 are not on the curve due to
    the long lived radioactive decay of $^{176}$Lu and $^{187}$Re.
    }
\end{figure}

Except for our $^{185}$W capture cross section, the other relevant stellar 
cross for the $s$-process calculations were taken 
from the compilation of \citet{Bao00}.
The $s$-process flow was normalized at the $s$-only isotope $^{150}$Sm
which represents an accurate measure of the unbranched $N\sigma$ curve
and calculated down to the W-Re-Os branching using programs described
by \citet{bcm97} with an average exposure of $\tau_0$=0.296 
($kT$/30\,keV)$^{1/2}$mbarn$^{-1}$.
This value corresponds also with a value reported by \citet{arla99}.
The relevant part of the calculation, the mass region from $A=145$ to
the termination of the $s$-process at $A=209$ is shown in 
Fig.~\ref{plotsix}. In this part the overall feature of the  
$N\sigma$ curve is a slow decrease up to $A=200$.    
This mass region includes not only the studied
W-Re-Os branching and the $s$-only $^{150}$Sm isotope on the unique
synthesis path used for normalization of the $N_{\rm s}\sigma_{\rm s}$
curve but also other important $s$-process branchings. These are first
the branchings that are sensitive exclusively to the neutron density,
i.e., the Nd-Pm-Sm ($A=147-150$), Er-Tm-Yb ($A=169-171$), W-Re-Os
($A=185-187$), and the Os-Ir-Pt ($A=191-193$) branchings, and second
the branchings dependent on neutron density, temperature and electron
density, i.e., the Sm-Eu-Gd ($A=151-152$), Eu-Gd ($A=154-156$), and
Dy-Er ($A=163-164$) branchings. To adjust the $s$-process neutron
density, temperature, and electron density, requires an $s$-only
isotope to be located inside the branching. These empirical data
points are shown in 
Fig.~\ref{plotsix} as full solid circles with smaller $N_{\rm 
s}\sigma_{\rm s}$ values than the $N_{\rm s}\sigma_{\rm s}$ value 
of the unique synthesis path represented by $N_{\rm s}\sigma_{\rm 
s}$($^{150}$Sm). For the 
W-Re-Os branching (Fig.~\ref{plotfive}) the branch point isotope is 
$^{186}$Os. %It should be noted that
%the $^{186}$Os $s$-process abundance consists of the directly 
%synthesized $^{186}$Os and the abundance synthesized as radiogenic $^{186}$Re.
%Similarly $^{148}$Sm contains radiogenic $^{148}$Pm, $^{154}$Gd
%radiogenic $^{154}$Eu, $^{170}$Yb radiogenic $^{170}$Tm, and $^{192}$Pt 
%radiogenic $^{192}$Ir. 

With the new stellar $^{185}$W cross section an average neutron density of
\begin{equation}
n_{\rm n} = 3.8 ^{+0.9} _{-0.8} \times 10^8\, {\rm cm}^{-3}
\end{equation}
was found. The uncertainty of the $^{185}$W cross section 
contributes by ($^{+16\,\%}_ {-12\,\%}$). The main uncertainty
in the present $n_{\rm n}$ determination comes from the 6.3\,\% 
uncertainty in the $N\sigma$ value of $^{186}$Os which transforms to an 
error of ($^{+23\,\%}_ {-20\,\%}$) in the neutron density. The 6.3\,\%  
uncertainty in $N\sigma$($^{186}$Os) is primarily the uncertainty of the 
solar osmium abundance \citep{andg89}. 

The present neutron density is still consistent with the Nd-Pm-Sm 
branching \citep{reif02} but is too high for reproducing the Er-Tm-Yb 
and - in particular - Os-Ir-Pt branchings. While the parameters 
of the first case are rather uncertain, the Os-Ir-Pt branching has recently
been studied with much improved cross sections yielding a 
neutron density of only $0.7 \times 10^8\, {\rm cm}^{-3}$
 \citep{koe01}. In Table~\ref{tablethree} our present value for the
 neutron density is compared with corresponding results from other
 branchings. 

The present neutron density in combination with the adopted values for the
 $s$-process temperature of 
$kT$=27.1\,keV \citep{bcm97}, and for the electron density of
$n_{\rm e} = 5.4 \times 10^{26}$\,cm$^{-3}$ \citep{arla99} provides also
to a fair reproduction of the other branchings
shown (Fig.~\ref{plotfive}). The empirical $N_{\rm s}\sigma_{\rm s}$ 
values of $^{152}$Gd and $^{164}$Er are underproduced in the 
calculation. This is reasonable as significant $p$-process 
contributions of up to 50\,\%  and about 10\,\% can be expected for 
$^{152}$Gd and $^{164}$Er, respectively.   

The discrepant neutron densities derived from different branchings as 
listed in Table~\ref{tablethree} indicate an inherent difficulty of the 
classical model due to the rather schematic assumption of constant
neutron density and temperature during the $s$-process. Hence, a
consistent description of the various branchings has to be based on more
realistic scenarios provided by  stellar model calculations. 

\subsection{The $s$-process in AGB stars}
\label{subsec:agb}
The main component of the $s$-process nucleosynthesis occurs during
helium shell burning in low mass AGB stars. The evolution of these
stars and the related $s$-process nucleosynthesis has been discussed
extensively by \citet{Gal98} and \citet{Bus01}, and it has been shown
that this model is able to reproduce the main $s$-process component
within 10\,\% \citep{arla99} as the result of the average composition
of the $s$-process abundance distribution of two AGB stellar models of
1.5 $M_{\odot}$ and 3 $M_{\odot}$ and a metallicity of $Z = 0.01$. 
In this model, the $s$-process is driven
by two neutron sources. The first, $^{13}$C$(\alpha,n)^{16}$O,
operates in the interpulse period between two helium flashes, and the
second, $^{22}$Ne$(\alpha,n)^{25}$Mg, is activated at higher
temperatures during the helium shell flash where almost the whole He
intershell, i.e.\ the region between the H shell and the He shell,
becomes convective for a relatively short period of time. The $^{13}$C
neutron source accounts for about 95\,\% of the total neutron exposure
in a thin radiative layer of about $10^{-4}\, M_{\odot}$;
however, the produced $s$-process abundances that depend on branching
points along the $s$-path are significantly
modified by the $^{22}$Ne source which is operating in the convective
helium burning zone. After a limited number of helium shell flashes,
at the quenching of the thermal instability, the convective envelope
penetrates in the top region of the He intershell, dredging up
$^{12}$C and $s$-process rich material. The envelope is progressively
lost by strong AGB winds and remixed into the interstellar medium.
%Finally, about one half of the produced
%$s$-process material is dredged up into the convective envelope and
%remixed into the interstellar medium by strong AGB winds.

It is important to emphasize that the profiles for neutron density and 
temperature are now provided by the stellar model. Therefore, the
abundance patterns of $s$-process branchings represent a critical test
for this stellar model.

The analysis of the branching at \WV\ shown in Fig.\ \ref{plotfive}
leads to a significant overproduction of the $s$-only isotope
$^{186}$Os by 20\,\% in this
stellar $s$-process model. All neutron capture cross sections have
been taken from the compilation of \citet{Bao00} with the exception of
the \WV \rng \WVI\ cross section where the present result was
used. This means that the model apparently overestimates the
$\beta^-$-decay part and/or underestimates the neutron capture part of
the \WV\ branching. Consequently, it underestimates the $s$-process
contribution to $^{187}$Re. Note that the uncertainties of the
previously existing calculated \WV\ cross sections  \citep{Bao00} were
generous enough to allow for a roughly consistent description of the
observed $^{186}$Os abundance.

The observed abundance of $^{186}$Os can be reproduced by the stellar model if
one increases the \WV \rng \WVI\ cross section by 60\,\%. However,
such an enhancement is outside the present experimental
uncertainties. If the \WV \rng \WVI\ cross section is
enhanced within the experimental uncertainties of about 20\,\%,
one still finds an overproduction of $^{186}$Os by 12\,\% which is
still slightly inconsistent with the observed abundance.

There are several possible explanations to cure this
problem. First, the $^{186}$Os$(n,\gamma)^{187}$Os cross
section could be 20\,\% larger than the adopted value by
\citet{Bao00}. However, this value is based on two independent
experiments who quote uncertainties between five and ten per cent
\citep{Winters82,Browne81}. Secondly, the additional branching at
$^{186}$Re could reduce the $^{186}$Os abundance either by an increased
$^{186}$Re$(n,\gamma)^{187}$Re cross section or by enhanced electron
capture of $^{186}$Re under stellar conditions. However, it has been
pointed out by \citet{Taka87} that under any realistic assumptions for
the capture cross section and the ratio between $\beta^-$-decay and
electron capture, the $^{186}$Re $\beta^-$-decay is always faster than
the neutron capture.

All these nuclear physics questions will be studied in additional 
experiments in the near future. Neutron capture experiments on osmium
isotopes are planned at the new n$\_$TOF facility at CERN and at the Karlsruhe
Van de Graaff accelerator, where uncertainties of the order of 1\,\% can be 
achieved using the $4\pi$ BaF$_2$ detector. Additionally, the
$^{187}$Re$(\gamma,n)^{186}$Re cross section shall be measured using
the monochromatic photon beam available from Laser-Compton
backscattering at AIST, Tsukuba, thus allowing to improve the cross
section for the inverse $^{186}$Re$(n,\gamma)^{187}$Re reaction. 
%Finally, the solar abundance of $^{186}$Os has a relatively big
%uncertainty of about 7\,\%.

With these improvements and by using a realistic $s$-process model
there is a good chance for analyzing the $^{185}$W and $^{186}$Re branchings 
with sufficient confidence to establish the abundance of $^{186}$Os as a 
sensitive test for the stellar model.

\section{Summary and Conclusions}
\label{sec:conc}
We have measured the photodisintegration cross section of the \WVI
\rgn\ \WV\ reaction at energies near the reaction
threshold. The experimental data have been used to restrict model
predictions for the $A = 186$ system and to derive the neutron capture
cross section of the inverse \WV \rng \WVI\ reaction. The result of
$\sigma =  (687 \pm 110)$\,mbarn is close to the calculated cross section
which was recommended by \citet{Bao00}, but exhibits significantly
improved reliability. %better accuracy.

The $s$-process flow at the branch point isotope \WV\ has been analyzed within
the classical $s$-process model and within a realistic stellar model
for AGB stars. With the classical model one obtains a neutron
density of $3.8 \times 10^8$/cm$^3$ compatible with the analyses of the 
branchings at $A=147/148$, but incompatible with the branchings at
$A=169/170$ and 191/192. This inconsistency indicates that the
assumptions of the classical model are too schematic to account for
the stellar situation, where the $s$ process takes place. The
corresponding analysis based on a more realistic stellar model
overestimates the $^{186}$Os abundance by 20\,\%. Presently, we are facing
the question whether this mismatch is related with remaining
uncertainties in other nuclear physics data or whether it originates
from the $s$-process model itself. If the nuclear physics
uncertainties can be further reduced, the $s$-process branching at
\WV\ can be interpreted as a sensitive test of models for the
important AGB phase of stellar evolution.

\acknowledgments
We thank the S-DALINAC group around H.-D.\ Gr\"af for the stable beam
during this activation experiment. Encouraging discussions with A.\
Richter and H.\ Utsunomiya are gratefully acknowledged. We thank 
the referee R.\ D.\ Hoffman for his encouraging report and helpful
comments. This work was
supported by the Deutsche Forschungsgemeinschaft (contracts
Zi\,510/2-1 and FOR\,272/2-2) and by the Italian MURST-Cofin 2000
Project ``Stellar Observables of Cosmological Relevance''.



\clearpage

%%%%%%%%%% Figures %%%%%%%%%%%%%%%%%%%%%%%%%%%%%%%%%%%%%%%%%%%%%%%%%%%%%%%%%%

%\clearpage

%\clearpage

%\clearpage

%\clearpage

%\clearpage

\clearpage

%%%%%%%%%% Tables %%%%%%%%%%%%%%%%%%%%%%%%%%%%%%%%%%%%%%%%%%%%%%%%%%%%%%%%%%%

\begin{table}
\begin{center}
\caption{
Properties of the tungsten and the gold target which were used during
the activation experiment.
\label{tableone}
}

\begin{tabular}{ccccc}
\tableline\tableline
Target          & Mass (mg)     & Nucleus       & Abundance (\%)  
        & $S_{\rm{n}}$ (keV) \\
\tableline
tungsten        & 5847.8(10)    & \WVI          &  28.6 & 7194 \\
gold            & 163.5(5)      & $^{197}$Au    & 100.0 & 8071 \\
\tableline
\end{tabular}
\end{center}
\end{table}

%\clearpage
\begin{table}
\begin{center}
\caption{
Decay properties of the residual nuclei \WV\ and $^{196}$Au.
\label{tabletwo}
}

\begin{tabular}{cccc}
\tableline\tableline
Nucleus & $T_{1/2}$ (d) & $E_\gamma$ (keV)      & $I_\gamma$ (\%) \\
\tableline
\WV             &  75.1(3)      & 125.4 & 0.0192(7)     \\
$^{196}$Au      & 6.1669(6)     & 333.0 & 22.9(6)       \\
                &               & 355.7 & 87.0(8)       \\
                &               & 426.1 & 6.6(8)        \\
\tableline
\end{tabular}
\end{center}
\end{table}

%\clearpage
\begin{table}
\begin{center}
\caption{
Neutron densities in the classical $s$-process model from various branchings.
\label{tablethree}
}

\begin{tabular}{cccl}
  \tableline\tableline
  \rule[0mm]{0mm}{3.75mm}
  Branch points & $s$-only isotope & $n_{n}$ ($10^8$\,cm$^{-3}$) 
  & Reference \\
  \tableline
  \rule[0mm]{0mm}{3.75mm}
  $^{95}$Zr    & $^{96}$Mo   &
  4~$^{+ 3}_{- 2}$ & 1 \\
  \rule[0mm]{0mm}{3.75mm}
  $^{147}$Nd/$^{147}$Pm/$^{148}$Pm   & $^{148}$Sm &
  3.0~$\pm$~1.1 & 1 \\
  & & 4.9~$^{+0.6}_{-0.5}$ & 2\\
  \rule[0mm]{0mm}{3.75mm}
  $^{169}$Er/$^{170}$Tm   & $^{170}$Yb &                   
  1.8~$^{+ 4.5}_{-0.8}$ & 1 \\
  \rule[0mm]{0mm}{3.75mm}
  $^{185}$W/$^{186}$Re    & $^{186}$Os      &
  4.1~$^{+ 1.2}_{-1.1}$ & 3 \\
  \rule[0mm]{0mm}{3.75mm}
  {\bf $^{185}$W/$^{186}$Re}    & {\bf $^{186}$Os} & 
  {\bf 3.8$ ^{\bf +0.9} _{\bf -0.8}$} & {\bf 4} \\
  \rule[0mm]{0mm}{3.75mm}
  $^{191}$Os/$^{192}$Ir    & $^{192}$Pt      &
  0.7~$^{+ 0.05}_{-0.02}$ & 5 \\
  
  \tableline
\end{tabular}
\tablerefs{(1) \citet{Kaepp90}; (2) \citet{reif02}; (3)
  \citet{Kaepp91}; (4) this work; (5) \citet{koe01}.}
\end{center}
\end{table}

\end{document}